\documentclass[draft]{aipproc}

\layoutstyle{6x9}

\begin{document}

\title{Gluon Vortices and Induced Magnetic Field in Compact Stars}



\author{Efrain J. Ferrer}{
  address={Department of Physics, Western Illinois University,
Macomb, IL 61455, USA} }

\begin{abstract}
 The natural candidates for the realization of color
superconductivity are the extremely dense cores of compact stars,
many of which have very large magnetic fields, especially the
so-called magnetars. In this paper we discuss how a color
superconducting core can serve to generate and enhance the stellar
magnetic field without appealing to a magnetohydrodynamic dynamo
mechanism.
\end{abstract}

\maketitle


\section{INTRODUCTION}

  Neutron stars created in the aftermath of supernova explosions
  are so compact that gravity, the weakest force in nature, binds
  nucleons there 10 times more strongly than the strong-nuclear force
  binds nucleons in nuclei. At the tremendous densities reached
  in the cores of those compact objects one expects the realization of a quark
  deconfined phase that produces a color superconducting state
  \cite{CS, CFL}.

  A common characteristic of neutron stars is their strong
  magnetization. Their surface magnetic fields range from
  $B=1.7\times 10^{8} G$ (PSR B1957+20) up to $2.1\times 10^{13} G$
  (PSR B0154+61), with a typical value of $10^{12} G$ \cite{Taylor}. There are observational discoveries of
  even strongly magnetized stellar objects- known as magnetars- with
  surface magnetic fields of order $B\sim 10^{14}-10^{16} G$ \cite{magnetars}. In the core of these compact objects, the field may be
  considerably larger due to flux conservation during the core
  collapse. By applying the scalar virial theorem it can be shown
  that the interior field can reach values of order $B \sim 10^{18}
  G$ \cite{Lai}.

  The observed stellar magnetic fields are supposed to be created by a magnetohydrodynamic dynamo mechanism that amplifies a seed
magnetic field due to a rapidly rotating protoneutron star. Thus,
the standard explanation of the origin of the magnetar's large
magnetic field implies that the rotation should have a spin period
$<3 ms$. Nevertheless, this mechanism cannot explain all the
features of the supernova remnants surrounding these objects
\cite{magnetar-criticism,Xu06}. Part of this rotational energy is
supposed to power the supernova through rapid magnetic braking,
implying that the supernova remnants associated with magnetars
should be an order of magnitude more energetic than typical
supernova remnants. However, recent calculations
\cite{magnetar-criticism} indicate that their energies are similar.
In addition, one would expect that when a magnetar spins down, the
rotational energy output should go into a magnetized particle wind
of ultrarelativistic electrons and positrons that radiate via
synchrotron emission across the electromagnetic spectrum.
Nevertheless, so far nobody has detected the expected luminous
pulsar wind nebulae around magnetars \cite{Safi-Harb}. On the other
hand, some magnetars emit repeated flares or bursts of energy in the
range of $10^{42} - 10^{46} erg$ \cite{burst}. Since the emitted
energy significantly exceeds the energy loss in their rotational
energy in the same period, it is natural to expect that the energy
unbalance could be supplied by the stellar magnetic field, which is
the only known additional energy source. However, from the spin
history of these objects, there is no clear evidence of the surface
magnetic field damping \cite{spin}.

Although more observations are needed to confirm the above issues,
current data indicate that alternative models to the standard
magnetar model \cite{magnetars} need to be considered. For example,
some authors \cite{magnetar-criticism, Soni} have suggested that
magnetars could be the outcome of stellar progenitors with highly
magnetized cores. In this paper I want to show that a progenitor
star with a color superconducting core would be capable of inducing
and/or enhancing its magnetic field through mechanisms different
than those of the standard magnetar \cite{magnetars}.

The color superconducting state of dense quark matter exhibits a
rich spectrum of different phases. For matter densities high enough
to guarantee the quark deconfinement, and sufficiently low to
consider the decoupling of the strange quark mass, $M_s$, the quarks
can condense to form a two-flavor color superconductor (2SC)
\cite{CS}. At moderately higher densities, for which $M_s$ cannot be
regarded extremely large, the three-flavor phases, $gCFL$
\cite{Alford} and $CFL$ \cite{CFL}, will appear respectively with
increasing densities. Matter in the stellar regions should be
neutral and remain in $\beta$ equilibrium. When these conditions are
taken into account in the $2SC$ phase \cite{Igor}, or when put
together with the $M_s$ contribution into the $gCFL$ phase
\cite{Fukushima}, some gluon modes become tachyonic, indicating that
the system ground state should be restructured to remove the created
chromomagnetic instability. Several interesting proposals have
already been worked out to find the stable ground state at the
realistic, moderate densities of compact stars. Some of the most
promising possibilities are a modified $CFL$-phase with a condensate
of kaons \cite{Schafer} that requires a space-dependent condensate
to remove the instability, a LOFF phase in which the quarks pair
with nonzero total momentum \cite{LOFF}, and homogenous
\cite{miransky}, as well as inhomogeneous \cite{Vortex}, gluon
condensate phases. At present it is not clear if any of these
proposals is the final solution to the problem; albeit, finding the
ground state that minimizes the system energy at densities of
interest for astrophysical applications is a pressing problem which
is attracting much attention.

In the following, I will discuss the mechanisms that can enhance
\cite{PCFL} and/or induce \cite{Vortex} magnetic fields in color
superconducting states. The realization of such mechanisms in the
core of compact stars can be an interesting alternative in order to
address the main criticisms pointed out above to the standard
magnetar's paradigm. On the other hand, having a mechanism that
associates the existence of high magnetic fields with color
superconductivity at moderate densities can serve to single out the
magnetars as the most probable astronomical objects for the
realization of this high-dense state of matter.

\section{ Magnetic-Field Increase by gluon vortices in color superconductors}

The interaction of an external magnetic field with dense quark
matter has been investigated by several authors \cite{Vortex} -
\cite{Phases}. Here I want to consider the effects of an external
magnetic field in the gluon dynamics of a color superconductor with
three massless quark flavors ($CFL$ phase) \cite{PCFL}, but the same
effect can be produced in the two-flavor theory ($2SC$ phase). The
ground state of the $CFL$ phase is formed by spin-zero Cooper pairs
in the color-antitriplet, flavor-antitriplet representation. An
important feature of spin-zero color superconductivity is that,
although the color condensate has non-zero electric charge, the
linear combination
$\widetilde{A}_{\mu}=\cos{\theta}\,A_{\mu}+\sin{\theta}\,G^{8}_{\mu}$
of the photon $A_{\mu}$ and the gluon $G^{8}_{\mu}$ remains massless
\cite{alf-berg-raj-NPB-02, alf-raj-wil-99/537}, so it behaves as an
in-medium electromagnetic field, while the orthogonal combination
$\widetilde{G}_{\mu}^8=-\sin{\theta}A_{\mu}+\cos{\theta}\,G^{8}_{\mu}$
is massive. In the CFL phase the mixing angle $\theta$ is small,
thus the penetrating "electromagnetic" field is mostly formed by the
original photon with only a small gluon admixture. Even though
gluons are neutral with respect to the conventional
electromagnetism, in the $CFL$ phase they are charged with respect
to the rotated field acquiring $\widetilde{Q}$ charges:
\begin{equation} \label{table}
\begin{tabular}{|c|c|c|c|c|c|c|c|c|}
  \hline
  $G_{\mu}^{1}$ & $G_{\mu}^{2}$ & $G_{\mu}^{3}$ & $G_{\mu}^{+}$ & $G_{\mu}^{-}$ & $I_{\mu}^{+}$ & $I_{\mu}^{-}$ & $\widetilde{G}_{\mu}^{8}$ \\
  \hline
  0 & 0 & 0 & 1 & -1 & 1 & -1 & 0 \\
  \hline
\end{tabular} \ ,
\end{equation}
given in units of $\widetilde{e} = e \cos{\theta}$. The
$\widetilde{Q}$-charged fields in (\ref{table}) correspond to the
combinations $G_{\mu}^{\pm}\equiv\frac{1}{\sqrt{2}}[G_{\mu}^{4}\mp
iG_{\mu}^{5}]$ and
$I_{\mu}^{\pm}\equiv\frac{1}{\sqrt{2}}[G_{\mu}^{6}\mp
iG_{\mu}^{7}]$. The external rotated magnetic field $\widetilde{H}$
that penetrates the CFL phase produces an instability when its
strength becomes larger than the Meissner mass $m_{M}$ of the
$\widetilde{Q}$-charged gluons. The tachyonic mode can be easily
found by diagonalizing the mass matrix of the field components
($G^{\pm}_{1}, G^{\pm}_{2}$)
\begin{equation}
\left(
\begin{array}{cc}
m_{M}^{2}& i\widetilde{e}\widetilde{H} \\
- i\widetilde{e}\widetilde{H}& m_{M}^{2}
 \label{mass-matrx}
\end{array} \right) \rightarrow
\left(
\begin{array}{cc}
m_{M}^{2}+\widetilde{e}\widetilde{H}& 0 \\
0& m_{M}^{2}-\widetilde{e}\widetilde{H}
 \label{mass-matrx}
\end{array} \right)
\end{equation}
Clearly, above the critical field $\widetilde{e}\widetilde{H}_{C}=
m_{M}^2$ the lowest mass mode in (\ref{mass-matrx}) becomes
tachyonic, with corresponding eigenvector of amplitude $G$ in the
$(1,i)$ direction for $G^{-}$ ($G^{\ast}$ in the $(1,-i)$ direction
for $G^{+}$). This phenomenon is the well known "zero-mode problem"
found in the presence of a magnetic field for Yang-Mills fields
\cite{zero-mode}, for the $W^{\pm}_{\mu}$ bosons in the electroweak
theory \cite{Skalozub, Olesen}, and even for higher-spin fields in
the context of string theory \cite{porrati}. To remove the tachyonic
mode, the ground state is restructured through the formation of a
gluon condensate of amplitude $G$, as well as an induced magnetic
field $\widetilde{B}$ that originates due to the backreaction of the
G condensate on the rotated electromagnetic field. The condensate
solutions can be found by minimizing with respect to $G$ and
$\widetilde{B}$ the Gibbs free-energy density
$\mathcal{G}_{c}=\mathcal{F}-\widetilde{H}\widetilde{B}$,
($\mathcal{F}$ is the free energy density)
\begin{eqnarray}
\label{Gibbs} \mathcal{G}_{c} =\mathcal{F}_{n0}
-2G^{\dag}\widetilde{\Pi}^{2}
G-2(2\widetilde{e}\widetilde{B}-m_{M}^{2})|G|^{2}+2g^{2}|G|^{4} +
\frac{1}{2}\widetilde{B}^{2}-\widetilde{H}\widetilde{B}
\end{eqnarray}
In (\ref{Gibbs}) $\mathcal{F}_{n0}$ is the system free energy
density in the normal-CFL phase ($G=0$) at zero applied field. Using
(\ref{Gibbs}) the minimum equations at $\widetilde{H}\sim
\widetilde{H}_{C}$ for the condensate $G$ and induced field
$\widetilde{B}$ respectively are
\begin{equation}
\label{G-Eq} -\widetilde{\Pi}^{2}
G-(2\widetilde{e}\widetilde{B}-m_{M}^{2})G=0,
\end{equation}
\begin{equation}
\label{B-Eq} 2\widetilde{e} |G|^{2}-\widetilde{B}+\widetilde{H}=0
\end{equation}
Identifying $G$ with the complex order parameter, Eqs.
(\ref{G-Eq})-(\ref{B-Eq}) become analogous to the Ginzburg-Landau
equations for a conventional superconductor except for the
$\widetilde{B}$ contribution in the second term in (\ref{G-Eq}) and
the sign of the first term in (\ref{B-Eq}). The origin of both terms
can be traced back to the anomalous magnetic moment term
$4\widetilde{e}\widetilde{B}|G|^{2}$ in the Gibbs free energy
density (\ref{Gibbs}). Notice that because of the different sign in
the first term of (\ref{B-Eq}), contrary to what occurs in
conventional superconductivity, the resultant field $\widetilde{B}$
is stronger than the applied field $\widetilde{H}$. Thus, when a
gluon condensate develops, the magnetic field will be antiscreened
and the color superconductor will behave as a paramagnet. The
antiscreening of a magnetic field has been also found in the context
of the electroweak theory for magnetic fields $H \geq
M_{W}^{2}/e\sim 10^{24} G$ \cite{Olesen}. Just as in the electroweak
case, the antiscreening in the color superconductor is a direct
consequence of the asymptotic freedom of the underlying theory
\cite{Olesen}.

The explicit solution of (\ref{G-Eq})-(\ref{B-Eq}) can be found
following Abrikosov's approach \cite{Abrikosov} to type II metal
superconductivity for the limit situation when the applied field is
near the critical value $H_{c2}$. In our case one finds \cite{PCFL}
that the ground state is restructured, forming a vortex state
characterized by the condensation of charged gluons and the creation
of magnetic flux tubes. In the vortex state the magnetic field
outside the flux tubes is equal to the applied one, while inside the
tubes its strength increases by an amount that depends on the
amplitude of the gluon condensate. This non-linear paramagnetic
behavior of the color superconductor can be relevant to boosting the
star's magnetic field without appealing to a magnetohydrodynamic
dynamo mechanism.

\section{Induced magnetic field in a color superconductor at moderate densities}

Let us consider now the Meissner unstable region of the so-called
gapped $2SC$ phase \cite{Huang} (we expect that the obtained results
can be also realized in the Meissner unstable region of the
three-flavor theory). As is known \cite{Igor}, the gapped 2SC turned
out to be unstable for certain values of the baryonic chemical
potential, even at zero applied magnetic field, once the gauge
fields $\{G_{\mu}^{(1)}, G_{\mu}^{(2)}, G_{\mu}^{(3)}, K_{\mu},
K_{\mu}^{\dag}, \widetilde{G}^{8}_{\mu}, \widetilde{A}_{\mu}\}$ were
taken into consideration. Note that even though the gluons are
neutral with respect to the regular electric charge, in the 2SC
phase the gluon complex doublet $K_{\mu}$ has nonzero rotated charge
of magnitude $\widetilde{q}=\widetilde{e}/2$, with $\widetilde{e} =
e \cos{\theta}$. The tachyonic modes in this unstable phase are
associated only with those charged gluon fields. It is natural to
expect then that a new, stable ground state should incorporate the
condensation of the $K_{\mu}$ gluons. In general we should allow for
an inhomogeneous condensate, and taking into account that this kind
of solution may generate rotated electromagnetic currents, the
rotated magnetic field should be included in the general framework
of the condensation phenomenon, but now as an induced field created
in the absence of an applied field.

Borrowing from the experience gained in the case with external
magnetic field, we propose the following ansatz for the condensate
solution
\begin{eqnarray}  \label{condensate-1}
\langle K_{\mu}^{T}\rangle  \equiv (\overline{G}_{\mu},0)/\sqrt{2},
\quad \overline{G}_{\mu} \equiv \overline{G}(x,y)(0,1,-i,0),
\end{eqnarray}
where we take advantage of the $SU(2)_{c}$ symmetry to write the
$\langle K_{i} \rangle$-doublet with only one nonzero component.
Since in this ansatz the inhomogeneity of the gluon condensate is
taken in the $(x,y)$-plane, it follows that the corresponding
induced magnetic field will be aligned in the perpendicular
direction, i.e. along the z-axes, $\langle\widetilde{f}_{12}
\rangle=\widetilde{B}$. The part of the free energy density that
depends on the gauge-field condensates is given by
\begin{equation}
\label{free-energy} \mathcal{F}_{g} =
\frac{\widetilde{B}^{2}}{2}-2\overline{G}^{\ast}\widetilde{\Pi}^{2}
\overline{G}+2g^{2}|\overline{G}|^{4}
 -2[2\widetilde{q}\widetilde{B}+(\mu_{8}-\mu_{3})^2-m_{M}^{2}]|\overline{G}|^{2}
\end{equation}
where the "chemical potentials" \cite{Linde} $\mu_{8}$ and $\mu_{3}$
are condensates of the time components of gauge fields, $\mu_{8}=
(\sqrt{3}g/2)\langle G_{0}^{(8)}\rangle$ and $\mu_{3}= (g/2)\langle
G_{0}^{(3)}\rangle$. From the neutrality condition for the
$3^{rd}$-color charge it is found that $\mu_{3}=\mu_{8}$. The fact
that $\mu_{3}$ gets a finite value just after the critical point
$m^{2}_{M}-\mu^{2}_{8} = 0$ is an indication of a first-order phase
transition, but since $\mu_{8}$ is parametrically suppressed in the
gapped phase by the quark chemical potential \cite{Igor}, it will be
a weakly first-order phase transition. As follows, we will consider
that $\mu_{3}=\mu_{8}$ in (\ref{free-energy}), and work close to the
transition point $\delta \mu_{c}$, which is the point where
$m_{M}^{2}$ continuously changes sign to a negative value. In that
region, the minimum equations for the condensate $G^{\ast}$ and
induced field $\widetilde{B}$ are respectively given by
\cite{Vortex}
\begin{equation}
\label{G-Eq-2} \widetilde{\Pi}^{2}
\overline{G}+\widetilde{q}\widetilde{B}\overline{G} \simeq 0.
\end{equation}
\begin{equation}
\label{B-Eq-1} 2\widetilde{q} |\overline{G}|^{2}-\widetilde{B}\simeq
0
\end{equation}

The relative sign between the two terms in Eq. (\ref{B-Eq-1})
implies that for $|\overline{G}|\neq 0$ a magnetic field
$\widetilde{B}$ is induced. This effect has the same physical root
as the paramagnetism found in the last section. The explicit form of
the inhomogeneity can be found from (\ref{G-Eq-2}) in the domain
$r\ll\xi$ (where $\xi^{2}\equiv
(g^{2}/\widetilde{q}^{2}-1)/|m_{M}^{2}|$ is the characteristic
length) as \cite{Vortex}
\begin{equation}
\label{Amplitude}
|\overline{G}|^2\simeq\frac{1}{2\widetilde{q}^2\xi^2}-\frac{r^2}{4\widetilde{q}^2\xi^4}
\end{equation}
The solution for $\widetilde{B}$ is found substituting
(\ref{Amplitude}) back into (\ref{B-Eq}). The induced field
$\widetilde{B}$ is homogeneous in the $z$-direction and
inhomogeneous in the $(x,y)$-plane.

Finally, it should be highlighted that when the absolute value of
the magnetic mass becomes of order $m_{g}$ \cite{Igor}, the gluon
condensate could produce a magnetic field $\sim 10^{16}-10^{17} G$.
The possibility of generating a magnetic field of such a large
magnitude in the core of a compact star without relying on a
magneto-hydrodynamic effect can be an interesting alternative to
address the main criticism \cite{magnetar-criticism} of the
observational conundrum of the standard magnetar paradigm. Then, one
concludes that if color superconductivity is realized in the core of
compact stellar objects at such expected densities that a Meissner
unstable phase is attained, the theory of the origin of stellar
magnetization should consider the discussed mechanisms.

\begin{theacknowledgments}
  This work has been supported in part by DOE Nuclear Theory
grant DE-FG02-07ER41458.
\end{theacknowledgments}



\bibliographystyle{aipproc}   


\end{document}